\documentclass[pdflatex,sn-mathphys-num]{sn-jnl}

\usepackage{graphicx}
\usepackage{multirow}
\usepackage{amsmath,amssymb,amsfonts}
\usepackage{mathrsfs}
\usepackage[title]{appendix}
\usepackage{textcomp}
\usepackage{manyfoot}
\usepackage{booktabs}
\usepackage{listings}
\usepackage{rotating}
\usepackage{tabularx}
\newcommand{\etal}{et~al.}
\newcommand{\dd}{\mathrm{d}}
\newcommand{\Ord}{\mathrm{O}}
\newcommand{\mat}{\mathbf}
\renewcommand{\vec}{\mathbf}
\newcommand{\bigstrut}{\rule{0pt}{10pt}}
\newcommand{\upstrut}{\rule{0pt}{10pt}}
\newcommand{\downstrut}{\raise-5pt\hbox{\rule{0pt}{10pt}}}

\begin{document}

\title[Drug-disease networks and drug repurposing]{Drug-disease networks and drug repurposing}

\author[1]{\fnm{Austin} \sur{Polanco}}\email{polancoa@umich.edu}

\author*[1,2]{\fnm{M. E. J.} \sur{Newman}}\email{mejn@umich.edu}

\affil*[1]{\orgdiv{Department of Physics}, \orgname{University of Michigan}, \orgaddress{\city{Ann Arbor}, \state{Michigan}, \country{United States of America}}}

\affil[2]{\orgdiv{Center for the Study of Complex Systems}, \orgname{University of Michigan},
\orgaddress{\city{Ann Arbor}, \state{Michigan}, \country{United States of America}}}

\abstract{Repurposing existing drugs to treat new diseases is a cost-effective alternative to \textit{de novo} drug development, but there are millions of potential drug-disease combinations to be considered with only a small fraction being viable.  In silico predictions of drug-disease associations can be invaluable for reducing the size of the search space.  In this work we present a novel network of drugs and the diseases they treat, compiled using a combination of existing textual and machine-readable databases, natural-language processing tools, and hand curation, and analyze it using network-based link prediction methods to identify potential drug-disease combinations.  We measure the efficacy of these methods using cross-validation tests and find that several methods, particularly those based on graph embedding and network model fitting, achieve impressive prediction performance, significantly better than previous approaches, with area under the ROC curve above 0.95 and average precision almost a thousand times better than chance.}

\maketitle

\section{Introduction}
Drug repurposing, the practice of finding new uses for established medications, is a vital part of the pharmaceutical development landscape~\cite{OM12,JBRD20,Pushpakom19}.  A fundamental part of the repurposing process is the identification of promising candidate drugs, and a significant amount of effort has been invested in the development of computational and statistical methods for performing this task~\cite{Cheng12,WGAJ13,BLFKXT14,Liu15}.  In this paper, we approach the problem using tools from the burgeoning field of network science~\cite{Newman18c}, viewing it as a link prediction problem~\cite{LK07} on a network of drugs and the conditions they treat.

Networks provide a convenient mathematical representation for many systems with complex patterns of interactions, including a number that are of interest in pharmacological or broader biomedical contexts.  Examples include networks of drug interactions~\cite{GS13}, networks of drugs and their targets~\cite{WLLT18}, and networks of genes and the diseases they are implicated in~\cite{GCVCVB07}.  A range of applications of network methods to medical and pharmaceutical problems have been pursued in recent years~\cite{GCVCVB07, Hopkins08, Cheng12, WGAJ13, GS13, WLLT18, LGMVG17, RGF22}.

In this paper we do two things: first, we compile an extensive network data set of drugs and the diseases they treat, using a combination of existing data, both machine-readable and textual, computational natural language processing, and human curation and data cleaning.  The network represents a total of 2620 drugs and 1669 diseases.  It differs from previous drug-disease data sets in being larger and more complete, and in being based solely on explicit therapeutic drug-disease indications, avoiding the use of associations inferred indirectly from, for example, drug function, targets, or structure.

\begin{figure}
\begin{center}
\includegraphics[width=8cm]{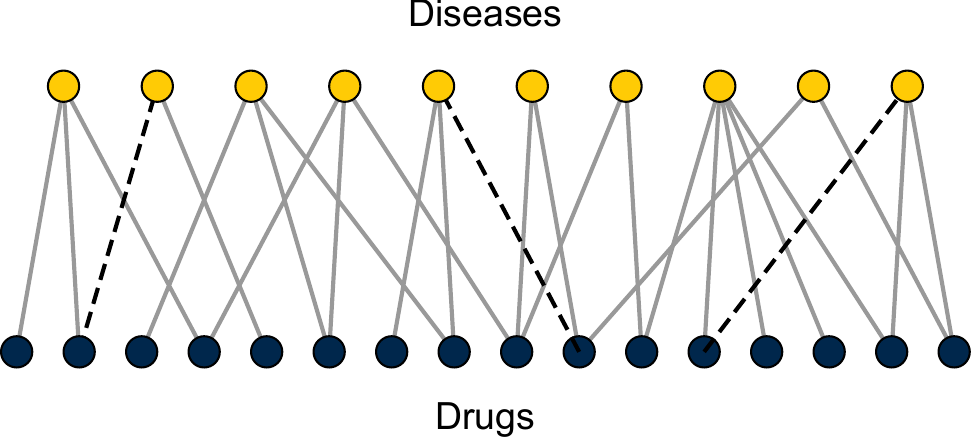}
\end{center}
\caption{A bipartite network of drugs and diseases of the type considered here.  Nodes in the network are of two types---drugs and diseases---and edges connect only nodes of unlike types to indicate which drugs are indicated for treatment of which diseases.  The network is assumed to be incomplete, so some edges that should be present are missing from the data (dashed lines).  Our goal is to identify these.}
\label{fig:bipartite}
\end{figure}

Second, armed with our data set we apply network methods to predict potential new therapeutic drug-disease pairs.  The data set takes the form of a \textit{bipartite network}, also known as an affiliation network, having two types of nodes, representing drugs and diseases, and connections only between unlike kinds: a connection, or edge, between a drug node and a disease node means that the drug is indicated for treatment of the disease---see Fig~\ref{fig:bipartite}.  Using this network, we identify candidates for drug repurposing by \textit{link prediction}.  It is a common finding in networks of all kinds that the available data are incomplete, that there are edges between nodes that should be present in the data but are not, either because of measurement error or because they have never been measured at all.  Link prediction~\cite{LK07,CMN08} is the process of attempting to identify these missing edges, usually based on observed patterns or regularities in the network.  For example, when examining a social network of friendships between individuals, it would be surprising if the data showed that two individuals with a lot of common friends were not themselves acquainted, so one might hypothesize that the data are incomplete---that this connection really does exist but has failed to be recorded for some reason.  It is a ``missing edge.''

A wide range of computational methods for identifying missing edges have been developed.  The simplest are little more than a formalized version of the ``lots of friends in common'' argument, but there are many more sophisticated ones as well.  One popular set of approaches makes use of graph representation learning methods such as non-negative matrix factorization~\cite{LS99}, node2vec~\cite{GL16}, and DeepWalk~\cite{PAS14}, which construct a low-dimensional embedding of the network.  Another promising approach makes use of statistical models of network structure, fitted to the input data.  The most commonly used model in this context is the degree-corrected stochastic block model~\cite{KN11a,Peixoto17}, although other models, such as hierarchical models~\cite{CMN08} and non-degree-corrected models~\cite{GS09} have also been tried.  Musawi~\etal~\cite{MRG23} have given a comprehensive review of link prediction methods as applied to biological networks.

Building on these works, in this paper we take a selection of link prediction algorithms, including both new and existing methods, and apply them to our network of drug-disease interactions.  We run extensive cross-validation tests to quantify the performance of each algorithm, removing a small fraction of edges at random from the network and testing the algorithm's ability to tell which ones were removed.

Our primary finding is that a subset of the algorithms perform well on this task, showing an impressive ability to pinpoint the missing edges in the drug-disease network, as quantified by standard measures.  The best of the methods we consider can achieve a measured area under the ROC curve in excess of 0.95 and average prediction precision almost a thousand times better than chance (although not in the same algorithm).  Moreover, this performance is achieved using purely network-based methods---no pharmacological input, other than the network itself, is used.  It is reasonable to suppose that a combination of network-based prediction and pharmacological insight could improve the performance still further and our methods could thus be used either alone or as part of a hybrid prediction strategy.  Our goal here, however, is not to create such a hybrid strategy, but specifically to test the use of link prediction as a tool.  We regard our work as a proof of concept demonstrating the performance that can be achieved with link prediction in this context.

The remainder of this paper is organized as follows.  In Section~\ref{sec:previous} we discuss previous work on network-based methods for prediction of drug-disease therapeutic interactions, then in Section~\ref{sec:network} we introduce our dataset and describe the methods used to assemble it.  In Section~\ref{sec:prediction} we describe the link prediction algorithms for bipartite networks that we employ, some of which are previously published while others are new but similar to previous methods for unipartite networks.  Section~\ref{sec:results} presents the results of our cross-validation tests and a comprehensive set of measurements of algorithm performance.  We also report some formal results on link prediction performance that allow us to place bounds on the number of missing edges in the network and hence tell us about the quality of the data set and the number of potential opportunities for drug repurposing.  In Section~\ref{sec:concs} we give our conclusions.  Some technical details are presented in appendices.

\subsection{Previous work}
\label{sec:previous}
A number of authors have considered network methods for drug repurposing~\cite{GSRS11,WGWCL15,HYS13,ZYLWLHL18,AADNYCCSH21,CHTKIWWCF23}, although they have taken somewhat different approaches to the one presented here.  Gottlieb~\etal~\cite{GSRS11} assembled a network similar to ours but smaller---about a quarter of the size---using some of the same resources but different methodology.  They perform link prediction using similarity-based methods akin to the methods we describe in Section~\ref{sec:similarity}, and find moderately good performance, as we also do, though not competitive with the more sophisticated machine learning algorithms we consider.  Wang~\etal~\cite{WGWCL15} took a somewhat similar approach with a small study based on an early version of the DrugBank database and employing a collaborative filtering algorithm that exploits network projections.  Although they make only a small number of predictions, their work clearly shows the promise of these types of techniques.  Huang~\etal~\cite{HYS13} offer a good example of indirect inference of drug-disease interactions, combining measures of disease similarity derived from text mining with measures of drug similarity from chemical and protein-protein interaction data, and predicting drug-disease combinations using a network label propagation algorithm.  They obtain a number of medically relevant predictions, although overall measures of prediction performance are quite low.  Zhang~\etal~\cite{ZYLWLHL18} studied previously published datasets of known drug-disease associations along with pharmacological data such as structure, targets, and drug interactions, and proposed a new algorithm that combines these elements to predict drug-disease interactions.  The associations in the network include not only therapeutic interactions but also other drug-disease associations such as side effects, and hence the predictions also include non-therapeutic associations, a potentially useful output, although different from our work, in the which the goal is to predict therapeutic associations only.  Cohen~\etal~\cite{CHTKIWWCF23} performed an unusual study that aimed to predict treatments for one specific disease, Covid-19.  This makes their network highly imbalanced: it has 8070 drug nodes but only 33 disease nodes, all versions of Covid-19.  Prediction in this setting is a different task from the one we consider, but Cohen~\etal\ achieved some promising results using a neural network method.

Abbas~\etal~\cite{AADNYCCSH21} applied a range of link prediction methods, including some of the same ones we consider, to various pharmacological networks, including drug-target and drug-interaction networks as well as a drug-disease network.  Their prediction results for the drug-disease network are good, particularly in terms of average precision, but come with a caveat: based on the sheer number of drug-disease interactions they claim, it appears unlikely that all of these interactions correspond to confirmed therapeutic uses of drugs.  Although details about the data set are scarce, it appears that the majority of edges in the network are indirectly inferred from other data.  Moreover, the much larger number of interactions means that the network density is significantly higher overall, which artificially increases algorithm precision under cross-validation---it is easier to make correct predictions if there are more such predictions to be made.

\begin{table}
\caption{Comparison of the network studied in this paper with networks of drugs and diseases compiled in previous work.  The final column indicates the type of data used to construct the network and/or make predictions: direct therapeutic interactions~(Th), structure~(S), genes~(G), targets~(Ta), enzymes~(E), pathways~(P), drug-drug interactions~(DD), disease ontology~(O).  Missing data are denoted by ``--''.}
\label{tab:previous}
\begin{tabular}{lrrrl}
   & Drugs & Diseases & Interactions & Basis \\
\hline
This paper & 2620 & 1669 & 8946 & Th \\
Gottlieb et al.~\cite{GSRS11} & 593 & 313 & 1933 & Th \\
Wang et al.~\cite{WGWCL15} & 963 & 1263 & -- & G, O \\
Zhang et al.~\cite{ZYLWLHL18} & 269 & 598 & 18\,416 & Th, S, Ta, E, P, DD \\
Zhang et al.~\cite{ZYLWLHL18} & 1323 & 2834 & 49\,217 & Th \\
Abbas et al.~\cite{AADNYCCSH21} & 5535 & 1662 & 466\,656 & -- \\
\end{tabular}
\end{table}

Table~\ref{tab:previous} compares our data set with the data sets used in previous studies.  Some, such as the smaller data set of Zhang~\etal~\cite{ZYLWLHL18}, are of relatively modest size.  Others such as the data set of Abbas~\etal~\cite{AADNYCCSH21} are much larger, but also more speculative as mentioned above.  We give a comparison of the performance of previous prediction methods with the method of this paper in Section~\ref{sec:concs}.

\section{Methods}
\label{sec:methods}
Our first task is the construction of the bipartite network of drugs and the diseases they treat, which involves a number of steps.

\subsection{Drug-disease network}
\label{sec:network}
As discussed in the introduction, our network is based solely on known therapeutic drug-disease associations.  The starting point is the DrugBank database (version 5.1.10, circa 2024)~\cite{drugBank17}, an online index of over 15\,000 drugs, with targets, chemical data, prescribing information, and other details.  Many of these drugs are experimental or of dubious therapeutic value and we remove a significant number from the set, including drugs not approved for clinical use, drugs labeled as supplements, cosmetics, food or food additives, household products, allergens, or contrast agents, and drugs belonging to no known category.  A full description of which drugs are removed is given in Appendix~\ref{app:data}.

Absent from the DrugBank database is a concise list of the conditions each drug is used to treat.  To obtain such a list we use a combination of strategies.  Some drugs in the database are accompanied by a Unique Ingredient Identifier or UNII code, an alphanumeric code that allows cross-listing with other databases.  Using these identifiers we queried another database, the NIH NCATS Inxight database~\cite{ncats22}, which includes machine-readable information on drug indications.  DrugBank entries without a UNII code, or for which no uses are listed on Inxight, were further divided into two groups, those for which DrugBank contains human-readable indications, and those for which it does not.  Drugs in the latter category were queried on the DailyMed database, another, smaller US federal database, which contains human-readable indications for some drugs.  If no indications are found the drug was discarded.

This leaves us with a subset of drugs with human-readable indications, either from DrugBank or from Daily\-Med, typically in the form of a few sentences of English text.  These indications were parsed using the Open\-AI large language model (LLM) GPT-3.5 to return a machine-readable list of diseases and conditions for each drug. (We also conducted some tests with GPT-4, the most advanced model available to us at the time of this study, but found no discernable difference in performance that would justify using this more expensive model.) Some drug indications contained no disease information that the LLM was able to extract and these were discarded. (A subset of the drugs for which no interactions were found were also checked manually and no instances of overlooked interactions were found.)  Finally, the entire network was reviewed by hand to confirm its accuracy. This included verifying that each interaction extracted by the LLM was genuinely supported by the corresponding drug indication, in order to catch potential LLM ``hallucinations,'' and the consolidation of duplicates---drugs or diseases that appear under multiple names.  For example, we found entries for ``Type II diabetes'' and ``Diabetes mellitus 2,'' which refer to the same condition.

\begin{figure}
\begin{center}
\includegraphics[width=10cm]{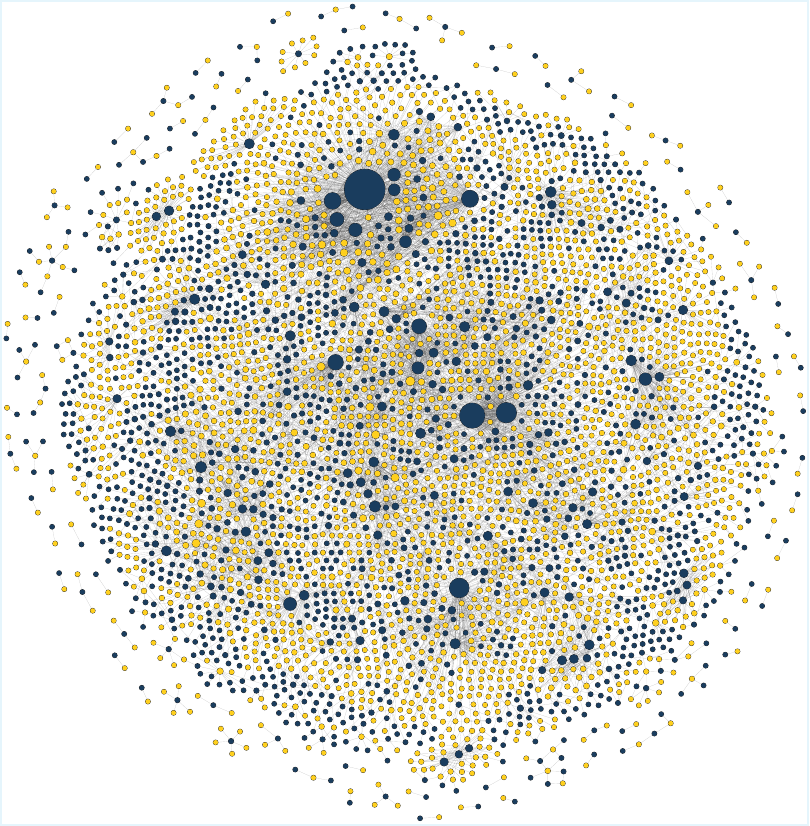}
\end{center}
\caption{Visualization of the complete network of drugs, diseases, and their therapeutic interactions.  Drug nodes are shown in blue and disease nodes in yellow.}
\label{fig:ddnet}
\end{figure}

\begin{figure}
\includegraphics[width=\textwidth]{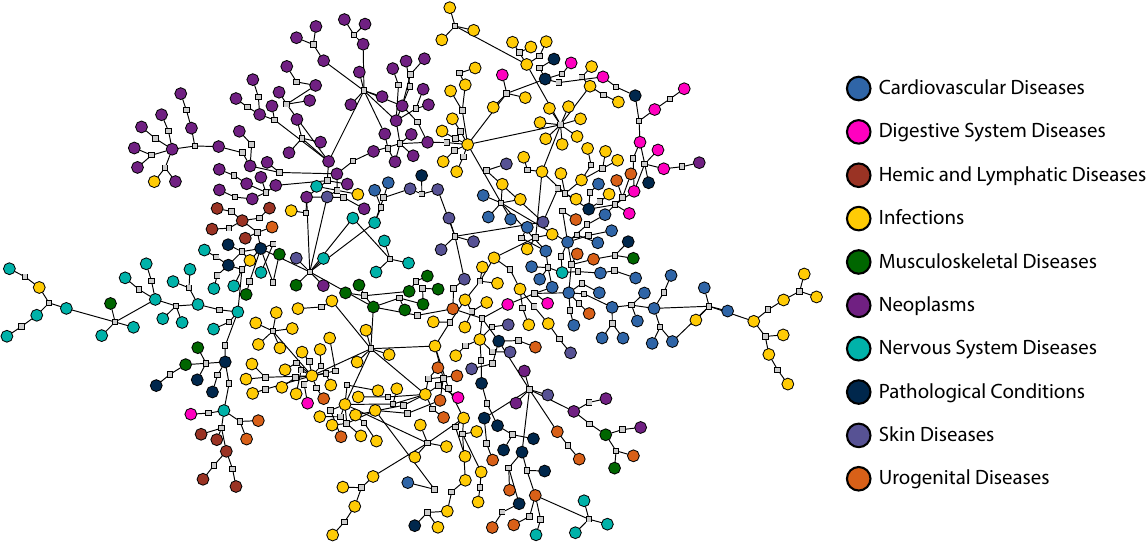}
\caption{A visualization of the network of drugs and diseases.  In this figure, we show (circles) a subset of the diseases in our network, those falling in the ten most common disease categories, as labeled, along with the shared drugs that connect them together (gray squares).}
\label{fig:disnet}
\end{figure}

The final data set consists of 2620 drugs and 1669 diseases, with 8946 edges connecting drugs to diseases they are known to treat.  This leaves over 4.3~million unconnected drug-disease pairs.  It is our goal to predict which among these are the most promising candidates for drug repurposing.  An image of the complete network is shown in Fig~\ref{fig:ddnet}, although the sheer number of nodes makes clear visualization challenging.  Fig~\ref{fig:disnet} shows perhaps a more informative visualization, of a portion of the network, corresponding to diseases in ten common disease categories and drugs that treat them.

\subsection{Link prediction}
\label{sec:prediction}
Link prediction~\cite{LK07} is the task of estimating which edges are missing from an observed network.  In the context of our drug-disease network this is equivalent to predicting specific drugs that can be used to treat specific diseases.  Link prediction is a well-studied problem in network science and a range of methods have been proposed for the task.  Our primary goal in this paper is to investigate the efficacy of these methods for drug repurposing.

Link prediction methods can in principle make use of a variety of ancillary information about network nodes and their characteristics.  For instance, one can imagine a method that employed chemical or structural information about drugs, clinical knowledge such as prescribing practice, or information about drug targets.  A number of techniques that use such information for identifying drug repurposing candidates have been proposed in the past~\cite{Cheng12,HYS13,ZYLWLHL18,AADNYCCSH21,Belyaeva21,BKBR21}.  Here, however, we take a purely network-driven approach that employs only the topological information contained in the network of drugs and diseases.

All network-based link prediction methods adopt basically the same framework.  One computes some ``prediction score'' for every pair of nodes, which indicates, often on an arbitrary scale, how likely those nodes are to be connected by an edge, with higher values denoting higher likelihood.  Then one sorts those values in decreasing order to create a list of potential missing edges from most to least likely.  Typically, interest will focus on the initial entries in the list, which represent the most promising candidates, and a good link prediction algorithm is one for which a large number of these initial entries turn out to be correct predictions.

The heart of the algorithm is the definition and calculation of the prediction scores.  Here we consider a variety of possible scores, drawing inspiration from various sources, including previously proposed algorithms for unipartite networks, representation learning methods, probabilistic models of network structure, and methods developed for the related problems of collaborative filtering and document classification.  In detail the methods we consider are as follows.

\subsection{Elementary algorithms}
\label{sec:similarity}
In some contexts successful link prediction can be performed using surprisingly simple heuristics.  Liben-Nowell and Kleinberg~\cite{LK07} have tested a range of such methods based on node degrees, path lengths, and network similarity measures.  Perhaps the simplest of these methods is the degree-based heuristic they call ``preferential attachment,'' after the well-known class of network growth processes by that name.  Under this method, the prediction score for a drug-disease node pair~$u,v$ is simply equal to the product $d_u d_v$ of the degrees (i.e.,~the number of edges) at each of the nodes.

Another group of elementary algorithms are those based on node similarity.  Similarity measures in network science are measures that quantify how similar pairs of nodes are in purely topological terms.  For unipartite networks the most common similarity measures, such as cosine similarity and Jaccard coefficient, are proportional to the number of network neighbors two nodes have in common, differing only in how that number is normalized.  One can then use these similarity measures directly as prediction scores.

Generalizing this notion to bipartite networks involves some subtleties.  We want to predict edges between nodes of unlike kinds (drugs and diseases in our application), but by definition such nodes do not share any neighbors: a~disease node only has drug neighbors and a drug node only has disease neighbors.  Instead, therefore, we define a prediction score between a drug node~$u$ and a disease node~$v$ to be the sum of the similarities between $v$ and other diseases that $u$ is known to treat.  In other words, our link prediction algorithm looks for diseases that are similar to those for which $u$ is already used.

In mathematical terms, we define~$\sigma(v,v')$ to be the similarity between disease nodes $v$ and~$v'$, and $N(u)$ to be the set of diseases that $u$ is known to treat.  Then the prediction score for drug~$u$ to treat disease~$v$ is
\begin{equation}
x(u,v) = \!\!\sum_{v'\in N(u)} \!\!\sigma(v,v').
\label{eq:xsim}
\end{equation}
One can also imagine defining prediction scores based on similarity between drugs: we could consider a particular disease~$v$ and look for drugs that are similar to ones currently used to treat it.  This would give the alternate definition
\begin{equation}
x(u,v) = \!\!\sum_{u'\in N(v)} \!\!\sigma(u,u').
\label{eq:xsim2}
\end{equation}
We have experimented with both approaches, but find that Eq.~\eqref{eq:xsim} gives distinctly superior performance to Eq.~\eqref{eq:xsim2}, so we do not pursue~\eqref{eq:xsim2} further here.

Equation~\eqref{eq:xsim} can be implemented with any of the many standard measures of node similarity in networks.  We here consider the five with definitions given in Table~\ref{tab:sim_form}: common neighbor count, cosine similarity, Jaccard coefficient, Dice (or Dice-S{\o}rensen) coefficient, and the hub-suppressed index.  All of these are based on the number of common neighbors between nodes, as described above, or equivalently the number of paths of length two.  In addition, we consider one further measure, which we call the ``Katz similarity'' in recognition of its close similarity to the well-known Katz centrality~\cite{Katz53}, which counts paths of all lengths, but with less weight given to longer ones~\cite{LHN06}.  All of these similarity measures were originally developed for use with standard (non-bipartite) networks, but they can be applied straightforwardly, without modification, to our bipartite case.  We have also experimented with some other common similarities, but find that they give clearly inferior results, so we do not pursue them.

\begin{table}
\caption{Definitions of similarity measures $\sigma(u,v)$ between network nodes used in Eq.~\eqref{eq:xsim}.  In these expressions, $u$ and $v$ are nodes, $d_u$ and $d_v$ are their degrees (the number of edges connected to them), and $n_{uv}$~is the number of common network neighbors between $u$ and~$v$.  For nodes of degree zero, all of these measures (except the common neighbors measure) give $0/0$, in which case the similarity is defined to be zero.}
\setlength{\tabcolsep}{8pt}
\begin{tabular}{lc}
Similarity measure     & Formula \\
\hline
Common neighbors\bigstrut     & $n_{uv}$ \\[2mm]
Cosine similarity & $\dfrac{n_{uv}}{\sqrt{d_u d_v}}$ \\[4mm]
Jaccard coefficient & $\dfrac{n_{uv}}{d_u+d_v-n_{uv}}$ \\[4mm]
Dice coefficient & $\dfrac{n_{uv}}{\frac12(d_u+d_v)}$\\[4mm]
Hub-suppressed index & $\dfrac{n_{uv}}{\max(d_u,d_v)}$
\end{tabular}
\label{tab:sim_form}
\end{table}

\subsection{Machine learning}
The algorithms of the previous section are simple heuristics that do not attempt to make use of any deeper structure in the network, but a variety of other algorithms have been proposed that employ machine learning methods to extract more complex structural information.  Here we consider a representative set of algorithms from this class, including standard algorithms and some perhaps less familiar recent proposals.

\textit{Singular value decomposition (SVD):} This standard matrix-based method constructs a low-rank approximation to the incidence matrix~$\mat{B}$ of the bipartite network.  The incidence matrix is the rectangular matrix with elements~$B_{uv}=1$ if drug~$u$ is indicated to treat disease~$v$ and 0 otherwise.  In this algorithm one first computes the singular value decomposition $\mat{B} = \mat{USV}^T$, where $\mat{U}$ and $\mat{V}$ are orthogonal matrices and $\mat{S}$ is the diagonal matrix of singular values, then one discards (i.e.,~sets to zero) all but the $K$ largest singular values, giving a modified diagonal matrix~$\mat{S}'$, and then computes the rank-$K$ matrix $\mat{B}' = \mat{US}'\mat{V}^T$.  The elements of this matrix are used to predict the missing edges---larger (more positive) elements indicate higher prediction certainty.  We use the SVD implementation in the LAPACK linear algebra library.  We have tested the algorithm for various values of~$K$ and find the best results around $K=60$.

\textit{Probabilistic latent semantic analysis (PLSA):} An important class of algorithms are those based on graph embeddings.  These methods attempt to place the nodes of a network at positions in a Euclidean space such that nodes with similar positions (in some sense) are connected by edges and others are not.  Perhaps the simplest version of this idea in the present bipartite context is one in which we assign vectors $\vec{r}_u$ to nodes of one type and $\vec{s}_v$ to nodes of the other type, and the edge between $u$ and $v$ is a random variable with probability equal to the inner product $\vec{r}_u\cdot\vec{s}_v$.  The dimension~$K$ of the vectors is a free parameter that can be tuned to give optimal results.  This approach has been employed particularly in document classification, leading to the method known as probabilistic latent semantic analysis or PLSA~\cite{Hofmann99}.  Here we retask this method for our drug-disease network and use an expectation-maximization (EM) algorithm to fit the PLSA model (see \cite{Hofmann04} and Appendix~\ref{app:plsa}), then use the probabilities $\vec{r}_u\cdot\vec{s}_v$ to predict the most likely missing edges.  We find best results for vector dimension around $K=90$.

\textit{Non-negative matrix factorization (NNMF):} The embedding used for PLSA can be thought of as an approximate decomposition of the incidence matrix into a product of two non-negative matrices whose columns are the vectors $\vec{r}_u$ and~$\vec{s}_v$.  In addition to the EM algorithm above, a~variety of other methods exist for finding such decompositions, based on various notions of approximation error, with the most common using a simple mean-squared error~\cite{LS99,LS01}.  Such approaches are known generically as non-negative matrix factorization algorithms.  We use the implementation in the \texttt{scikit-learn} Python package, which employs a mean-squared error.  We find best results for vector dimension around $K=80$.

\textit{Node2vec}: Node2vec is a graph embedding method developed by Grover and Leskovec~\cite{GL16} that uses biased random walks to train a machine learning model.  The random walks generate sequences which are used as neighbor sets for nodes, from which the model then learns an embedding.  The embedding dimension is a free parameter and we find best results for dimension $K=256$.  We use the implementation of node2vec in the \texttt{pytorch-geometric} Python package~\cite{FL19}.

\textit{Generic bipartite network embedding (GEBE\textsuperscript{p})}: Another embedding-based method is the GEBE\textsuperscript{p} method of Yang~\etal~\cite{YSHX22}, which finds embeddings by splitting the problem into two parts: first it computes a similarity between nodes of the same type, analogous to the measures of Section~\ref{sec:similarity}, then it computes weighted paths between nodes of different types.  These are then combined into a single objective function and optimized to learn the embedding.  Link prediction is performed by training a logistic regression classifier on the combined embedding vectors of node pairs and using the output of the classifier as the prediction score.

\textit{Bayesian personalized ranking (BPR):}
Bayesian personalized ranking~\cite{RFGS09} is a matrix factorization technique, originally developed for recommender systems, that works by maximizing the probability for each drug individually that a known interaction between the drug and a disease is ranked higher than an unknown one.  We use the implementation in the \texttt{LightFM} Python package~\cite{Kula15}.

\textit{Intra-class connection triadic closure (ICTC)}: The ICTC method of Shin~\etal~\cite{SGPKK20} leverages artificial neural networks to perform link prediction.  The method uses a linear graph autoencoder to learn implicit similarities between nodes of the same type, then uses these to predict the presence of edges in a manner similar to the methods of Section~\ref{sec:similarity}.

\subsection{Probabilistic network models}
An alternative approach to link prediction is a statistical one in which one fits a probabilistic model of network structure to the observed network.  Various models have been used for this purpose.  An early example is the work of Clauset~\etal~\cite{CMN08}, who proposed a hierarchical model that captures structure at multiple scales.  More recent work has found success using various versions of the stochastic block model (SBM)~\cite{HLL83}.  Guimer\`{a} and Sales-Pardo~\cite{GS09} were among the first to adopt this approach, employing the SBM in its original unmodified form, but better results are typically found using the variant ``degree-corrected'' SBM~\cite{KN11a,Peixoto17}.  In this model nodes are divided into some number of groups, and edges fall between them with probabilities that depend on group membership, but with a bias that increases probabilities for edges connected to nodes with high degree in the observed network.  Link prediction is performed by calculating the change in the likelihood of the network when a single edge is added and using these changes as the prediction scores.

We consider two variants of this approach.  The first is a Bayesian version of the standard degree-corrected model in which group memberships are sampled using a single-node Markov chain Monte Carlo algorithm~\cite{RCRN17} and prediction scores are averaged over the resulting samples.  The second is a ``microcanonical'' version of the model in which the numbers of edges, rather than their probabilities, depend on group membership, and states are sampled using a non-local cluster Monte Carlo~\cite{Peixoto17}.

\section{Results}
\label{sec:results}
We evaluate the performance of each of our link prediction algorithms on the drug-disease network using cross-validation.  We randomly remove 10\% of the edges in the network then measure the ability of an algorithm, when applied to the remaining network, to tell which ones were removed.  The procedure is repeated 50 times for each algorithm and the results averaged.  Success is measured by four standard metrics:
\begin{enumerate}
\item \textit{Area under the Receiver Operating Characteristic curve (AUROC):} Perhaps the most common measure of binary classifier performance, this measure is equal to the area under the ROC curve, i.e.,~the curve of true positive rate as a function of false positive rate, from start to finish of a run of the algorithm, all the way from the most promising prediction to the least.  Thus each complete run for each algorithm produces a single numerical AUROC value.  Values run from 0.5 to~1, with higher values being better.  AUROC can be thought of as a measure of how thorough or complete an algorithm's predictions are.  If we consider one of the removed edges to have been successfully predicted if an algorithm ranks that edge higher than the average non-edge, then the AUROC score is equal to the fraction of edges successfully predicted.  Thus a value close to 1 indicates an algorithm that gives very complete results and misses few of the predictions it aims to make.  The AUROC value can also be used to estimate the number of unknown interactions waiting to be discovered in a network---see Section~\ref{sec:bounds}.  A weakness of the measure is that AUROC values are insensitive to dilution of the predictions by false positives: there can be many wrong results among the right ones without affecting the score greatly, a problem that becomes particularly apparent in sparse data sets such as ours.
\item \textit{Area under the precision/recall curve (AUPR):} In practical situations, we often do care about dilution of the predictions with false positives, in which case the precision---the fraction of predictions that are correct---can be a more useful measure of performance.  AUPR averages the precision over values of the recall (i.e.,~the true positive rate), which auto\-matically weights the results towards the most promising predictions (those with highest prediction score), since recall varies most rapidly in the early part of a run.  We quote AUPR results as a percentage and they can be read, broadly speaking, as the average probability that a prediction the algorithm makes is correct.
\item \textit{Normalized AUPR:} For sparse data sets we expect the raw precision numbers to be small---we are searching for a small number of needles in a very large haystack, so even an algorithm that does far better than chance will still have low precision.  For this reason precision is often normalized relative to the baseline prevalence, i.e.,~the fraction of true positives in the entire test set.  The normalized value measures how much more likely the algorithm is to return a true positive on the average guess than would a random, no-skill classifier.
\item \textit{Top-$k$ precision:} For large prediction problems such as ours it is often the top predictions that are of most interest.  A drug developer cannot reasonably be expected to look through all four million drug-disease combinations in our data set, but a carefully curated selection of the most promising candidates could be very useful.  The top-$k$ precision is a measure of an algorithm's ability to generate a high-quality selection.  It is equal to the fraction of correct predictions among the top~$k$.  In this paper we quote figures for the top 100.
\end{enumerate}

The extent to which one cares about ROC curves versus precision/recall curves or top-$k$ precision depends on one's goals.  A clinical researcher (or a patient) with an interest in a particular condition or set of conditions might focus on high AUROC values, since they would want a thorough algorithm that is successful at finding all or most of the promising repurposing candidates, including those they care about, and misses very few.  A~drug developer, on the other hand, might focus on algorithms with high precision, since these would return the highest-quality predictions, and hence offer the best chance of finding drugs that can be usefully repurposed.  In general, we find that higher AUROC figures correspond to lower precision, and vice versa, so one can score well on one or the other but not both.

\begin{figure}
\begin{center}
\includegraphics[width=8.2cm]{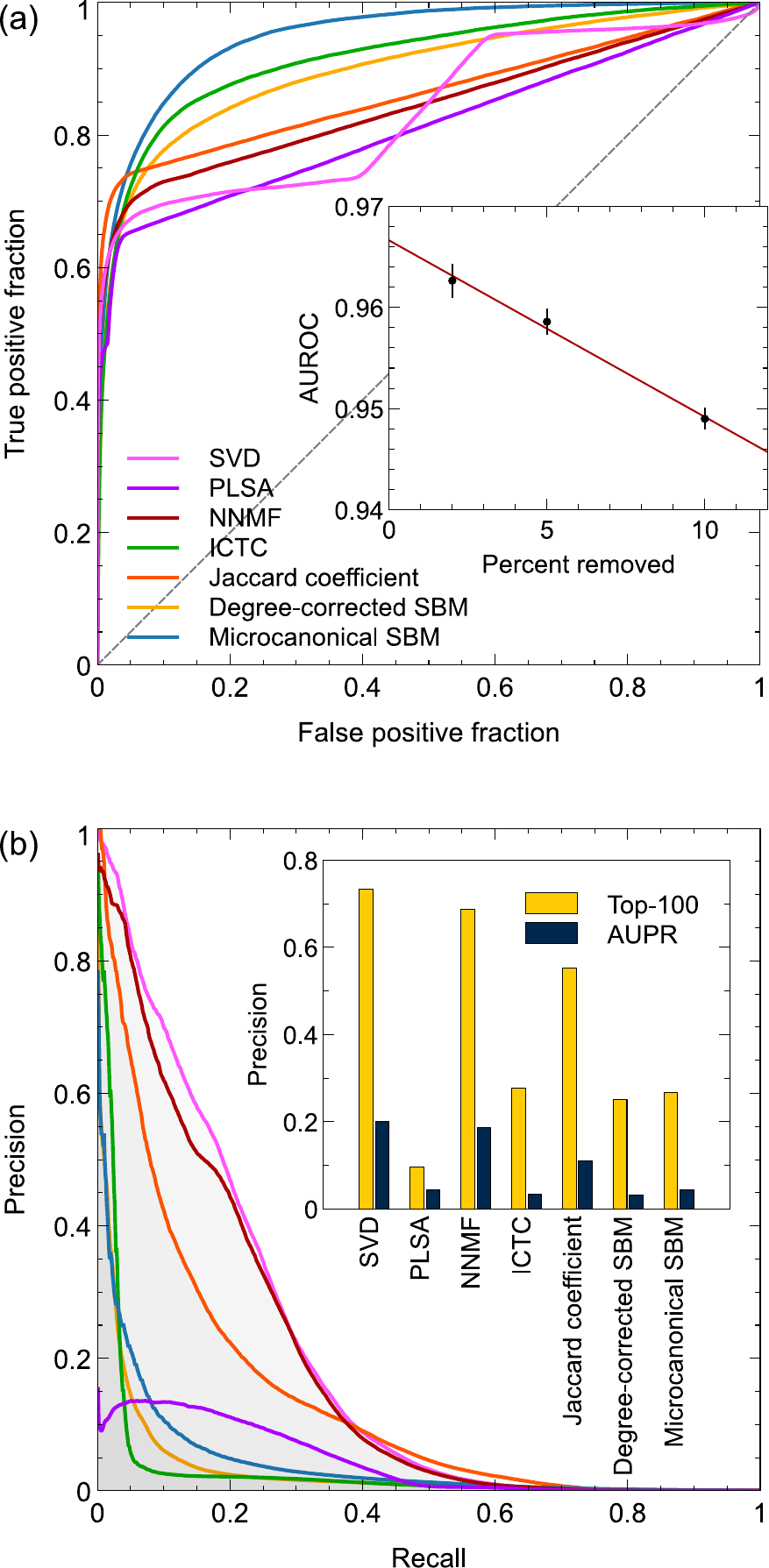}
\end{center}
\caption{(a)~Receiver operating characteristic (ROC) curves for seven of the best performing algorithms.  The dashed diagonal line represents the expected performance of a no-skill (random) classifier.  Inset: Area under the curve (AUROC) for the microcanonical SBM for various fractions of edges removed.  (b)~Precision/recall curves for the same selection of algorithms.  Colors are the same as in~(a).  Inset: Top-100 precision and area under the prediction/recall curve (AUPR) for each algorithm.}
\label{fig:curves}
\end{figure}

Fig~\ref{fig:curves} shows average ROC and precision/recall curves for a selection of our algorithms and Table~\ref{tab:lp} summarizes the values of the performance metrics listed above.  Inspecting the latter we note first that all algorithms give AUROC scores significantly above the baseline value of~0.5, indicating acceptable performance on the basic link prediction task.  There is nonetheless some significant variation.  The simple degree-based algorithm is the least competitive and not recommended for this application.  The similarity-based algorithms on the other hand do surprisingly well: they all have similar performance with AUROC scores around 0.86, except for the Katz similarity, which fares less well.  The machine learning algorithms PLSA, SVD, NNMF, node2vec, BPR, and GEBE$^\mathrm{p}$ are in the same vicinity but a little worse, but it is the final three algorithms that stand out for their impressive performance according to this metric---the ICTC deep learning algorithm and the two versions of the SBM, with the microcanonical SBM giving the best performance of all, with an AUROC score close to~0.95.

\begin{table}[b]
\caption{Performance measures for link prediction algorithms, estimated from 50 repetitions of cross-validation with 10\% of edges removed from the network.  We divide our algorithms into three categories: ``Elementary'' denotes algorithms based on node degrees or simple similarity measures such as counts of common neighbors between node pairs; ``Machine learning'' denotes methods such as matrix factorization, Bayesian, and deep-learning methods; ``SBM'' denotes network-based methods that make use of the stochastic block model.  The performance measures we use are the area under the ROC curve (AUROC), the area under the precision/recall curve (AUPR), area under the precision/recall curve normalized by prevalence, and precision over the top 100 predictions.  Numbers in parentheses indicate standard errors on the trailing digits.  Numbers in bold indicate the best performers.  Running time is for a single run of each algorithm.}
\footnotesize
\setlength{\tabcolsep}{3pt}
\begin{tabular}{lllrrrr}
&   & \multicolumn{3}{c}{Area under curve} \\
\cmidrule{3-5}
&   Algorithm & ROC    & Precision/recall & PR normalized & Top-100 precision & Time (sec) \\
    \hline
\begin{rotate}{90}
\hbox{\hspace{-5.3em}Elementary}
\end{rotate}
& Degree\upstrut & 0.721(2)  & 0.35(2)\% & 17(1) & 4.7(3)\% & 1 \\
& Common neighbors & 0.860(1) & 7.9(1)\% & 384(5) & 44.8(8)\% & 1 \\
& Cosine similarity & 0.860(1)  & 9.7(2)\% & 474(9) & 45.6(9)\% & 1 \\
& Jaccard coefficient & 0.863(1)  & 13.8(2)\% & 678(9) & 56.9(9)\% & 1 \\
& Dice coefficient & 0.862(1)  & 11.5(2)\% & 565(8) & 49.4(9)\% & 1 \\
& Hub-suppressed index & 0.861(1)  & 12.9(2)\% & 630(9) & 52.8(10)\% & 1 \\
& Katz similarity\downstrut~\cite{LHN06} & 0.789(2)  & 7.7(1)\% & 376(5) & 44.4(6)\% & 4 \\
    \hline
\begin{rotate}{90}
\hbox{\hspace{-6.5em}Machine learning}
\end{rotate}
& PLSA~\cite{Hofmann99,Hofmann04}\upstrut & 0.812(2) & 4.3(1)\% & 209(4) & 9.6(5)\% & 10 \\
& SVD & 0.835(1) & \textbf{20.0(2)\%} & \textbf{970(10)} & \textbf{73.3(7)\%} & 40 \\
& NNMF~\cite{LS99,LS01} & 0.844(1) & 18.6(2)\% & 908(11) & 68.7(7)\% & 57  \\
& Node2vec~\cite{GL16} &0.821(2)  &13.4(2)\% &655(10)  &51.0(9)\% &3984 \\
& GEBE\textsuperscript{p}~\cite{YSHX22} &0.750(1)  & 0.31(1)\% &15(1)  & 5.9(1)\% &20 \\
& BPR~\cite{RFGS09}   &0.860(1)   &9.98(16)\% &487(8)  &46.3(8)\% &17 \\
& ICTC\downstrut~\cite{SGPKK20} & 0.916(1) &3.3(1)\% & 161(3) &27.7(6)\% &242 \\
\hline
\begin{rotate}{90}
\hbox{\hspace{-1.2em}SBM}
\end{rotate}
& DCSBM\upstrut~\cite{GS09,KN11a,RCRN17}      & 0.898(1)  & 3.1(1)\% & 154(6) & 25.0(8)\% & 367 \\
& Microcanonical SBM~\cite{Peixoto17} & \textbf{0.949(1)} & 4.3(1)\% &209(5) &26.7(8)\% &4964    
\end{tabular}
\label{tab:lp}
\end{table}

Moreover, even this figure is an underestimate.  Under normal conditions the performance of any algorithm can be expected to improve as we increase the number of edges in the training data set, which we can do by removing fewer edges for cross-validation.  The inset to Fig~\ref{fig:curves} shows AUROC values for the microcanonical SBM for 10\%, 5\%, and 2\% of edges removed.  One cannot continue all the way zero---there have to be some edges to predict---but by extrapolating we estimate the AUROC at zero to be $0.967\pm0.001$.  Presumably in real-world applications of the method to drug repurposing one would use the entire data set, in order to get the best results possible, so this figure may be the most realistic one for practical applications.  (This assumes that one has the computational resources to analyze the entire network, but for a sparse network such as ours this is not an issue.)  Broadly speaking, the figure of 0.967 means that the microcanonical SBM algorithm is successful at identifying more than 96\% of true drug-disease interactions.

AUPR figures also vary substantially across algorithms, but they favor different algorithms from the AUROC scores.  All of the similarity-based methods return respectable AUPR values in the vicinity of 8\% or more.  The best is the algorithm based on the Jaccard coefficient, with an AUPR score of 13.8\% and a normalized value of 678, meaning that the algorithm is on average more than 600 times better at identifying drug-disease interactions than a no-skill random classifier.  The two versions of the SBM fare less well in this test and in particular the microcanonical model, which is so impressive in terms of the AUROC measure, scores only 4.3\%.  Among the machine learning algorithms some get very poor AUPR scores, such as the GEBE$^\mathrm{p}$ algorithm with a score under 1\%.  But the relatively simple singular value decomposition method is the standout on this test, with an AUPR score of 20.0\% and performance almost a thousand times better than chance.  Moreover, the leading predictions of this algorithm are substantially better even than this, with a top-100 precision value of 73.3\%, meaning that almost three-quarters of the first 100 predictions are correct.  The NNMF algorithm also does well, with a top-100 precision of 68.7\%.

From the point of view of pharmaceutical development these two algorithms, SVD and NNMF, may be the most promising.  Some of the others also perform well on the precision measures---the Jaccard, Dice, and hub-suppressed similarities all score around 50\% for instance---but are not competitive with SVD and NNMF.

Overall, these results suggest that network-based link prediction algorithms can be a useful tool for identifying candidates for drug repurposing, substantially reducing the amount of work necessary to make each successful identification.  (We do not give specific predictions of drug-disease pairings here, but interested readers can reproduce our entire set of over four million predictions from the posted data and code.)

In passing, we also note an interesting coincidence in the dimension of the representations of the network found by several of our algorithms.  For SVD we find that the ideal number of singular values to retain is about~60.  For NNMF the equivalent number is~80, for PLSA it is 90, and the degree-corrected stochastic block model finds about 30 communities each of drugs and diseases, for a total of around 60 overall.  (Exact numbers vary slightly during a run and from one run to another.)  The fact that these disparate algorithms are all successful in their predictions and give similar dimensions for the data may be a hint that there are about 60 to 90 different classes of drugs/diseases in the data, an observation that could be of pharmacological interest.  (An exception is the node2vec algorithm, for which the best results were obtained with a significantly higher embedding dimension of 256.)

\subsection{Running time}
\label{ssec:run_time}
The algorithms we consider vary substantially in the amount of time they take to run.  Approximate running times for a single run of each algorithm on all 4.3 million drug-disease combinations, measured in seconds of CPU time on conventional hardware circa 2024, are listed in Table~\ref{tab:lp}.  The elementary similarity-based methods, such as common neighbor counts and cosine similarity, are fastest, taking around one second of CPU time each.  (The Katz similarity is slower, at 4 seconds, but still fast.)  Although they are not the absolute winners in terms of prediction success, these algorithms do score well, particularly on precision, so they might be of use in cases where speed is important.

The machine learning algorithms are generally more computationally demanding, and in particular the singular value decomposition and non-negative matrix factorization methods are relatively slow, with SVD taking about 40 seconds of CPU time for a single run.  Even this, however, is not a significant amount of time in a typical application where one only needs to run the algorithm once.  Moreover, SVD and NNMF are both highly parallelizable and on modern multicore processors the wallclock running time of standard multithreaded implementations is only a fraction of the CPU time---15 seconds or so was typical in our tests of the SVD algorithm.

The three algorithms that perform best in terms of AUROC scores are also some of the most demanding---ICTC and the two versions of the stochastic block model.  The ICTC algorithm, like many neural network methods, can be accelerated by the use of GPUs, but without such aids takes 8 minutes per run.  And the microcanonical SBM, which has the best AUROC score overall, is the most demanding algorithm in our tests, with a running time of over an hour per run.

Balancing prediction success with speed, and assuming a preference for high precision rather than high AUROC scores, our overall pick for best algorithm is the singular value decomposition method, which runs in a few seconds and gives outstanding precision.  Non-negative matrix factorization is also competitive.  For applications requiring high speed, the similarity-based algorithms may be attractive, with the algorithm based on the Jaccard coefficient being the top performer in terms of prediction success.

\subsection{Bounds on the number of possible discoveries, false positives, and precision}
\label{sec:bounds}
In Appendix~\ref{app:bounds} we derive a theoretical bound on the AUROC statistic.  Under the assumption that our algorithms are equally good at predicting true missing edges in the network and edges that are randomly removed for cross-validation, we show that the value~$A$ of the statistic must satisfy $A \le 1 - \frac{1}{2}\mu$, where $\mu$ is the fraction of node pairs in the network that are not observed to be connected by an edge but which are in truth---these are edges that are missing from the data set and they represent the potential successful predictions that we could make using our algorithms.

Inverting the inequality, we see that
\begin{equation}
\mu \le 2(1-A),
\label{eq:mulimit}
\end{equation}
which places a limit on the number of missing edges waiting to be discovered in the network.  This inequality applies to the AUROC score for any algorithm and any fraction of edges removed for cross-validation, so we are at liberty to choose the algorithm that returns the highest value in order to achieve the best bound.  In our case, the highest value is the extrapolated value of $A=0.967$ obtained for the microcanonical SBM algorithm, and, substituting into Eq.~\eqref{eq:mulimit}, this implies that the fraction of possible drug-disease interactions remaining to be discovered in our network is at most $2\times(1-0.967) = 0.066$, or 6.6\% of the total.  At first glance this seems like a small fraction, but it still translates into a substantial number of potential predictions because of the sheer size of the network.  The number of node pairs unconnected by an observed edge is $2620\times1669 - 8946 = 4.36$ million and the maximum number of potential drug applications waiting to be discovered is 6.6\% of this figure, or about 288\,000, so there is plenty of room for exploration.

As further shown in Appendix~\ref{app:bounds}, we also have an upper bound $\nu \le 2(1-A)$ on the fraction~$\nu$ of false \emph{positives} in the data set, which is thus also limited to 6.6\%.  Because of the sparsity of the data set, however, this imposes a much sharper limit, since the number of observed interactions is relatively small, at just 8946.  Taking 6.6\% of this figure implies we have a maximum of just 590 possible false positives in our data set.  This tells us something about the quality of the data: at most 590 of the recorded therapeutic drug-disease combinations are in error.

Finally, as shown in Appendix~\ref{app:bounds}, the precision is also affected by the presence of missing edges in the network---it is reduced by a factor of $1-\mu$, as are measures proportional to precision such as AUPR and top-$k$ precision.  Given that $0\le\mu\le0.066$ in our case, we have $0.934\le1-\mu\le1$, which places relatively tight bounds on $1-\mu$.  In practice this means that the measured values of precision should be reasonably reliable and moreover that, to the extent they are modified in the presence of missing edges, they will be increased, not decreased, because the true precision is equal to the measured value divided by $1-\mu$.  Thus for instance the 20\% figure we find for AUPR under the SVD algorithm could be a (slight) underestimate---the true value could lie anywhere between 20.0\% and $20.0/0.934 = 21.4$\%.

\section{Discussion}
\label{sec:concs}
In this paper we have described the construction of a data set of 2620 drugs and 1669 diseases and conditions for which they are indicated, based on several pre-existing, publicly available databases, analyzed using a combination of machine learning methods and human data curation.  The resulting data set describes 8946 known drug-disease interactions.

We have used this data set to test the performance of a basket of algorithms for network link prediction, with the goal of identifying potential candidates for drug repurposing.  These methods regard the data set as a bipartite network of drugs and diseases, with edges in the network indicating which drug treats which disease, then attempt to identify potential missing edges, which would represent unknown drug-disease pairs.  This is a formidable task---there are more than four million possible pairs to consider---but nonetheless several of the algorithms appear to perform well.

Performance of each algorithm is tested using cross-validation, in which a small fraction (in this case 10\%) of edges are removed from the network and then the algorithm attempts to predict the removed edges.  We find good success with three algorithms in particular.  An algorithm based on the probabilistic network model known as the microcanonical stochastic block model returns an area under the ROC curve of 0.949, which increases to 0.967 when extrapolated to the full network without edges removed for cross-validation, meaning roughly speaking that the algorithm successfully predicts over 96\% of the missing edges in the network.  These predictions however may be diluted with false positives strewn among them, an issue that is addressed by another performance measure, the precision.  By this measure two other algorithms, based on singular value decomposition and non-negative matrix factorization, perform well.  These algorithms give AUPR scores of 0.20 and 0.19 respectively and precision of 73\% and 68\% on their top 100 predictions, indicating that over two-thirds of those predictions are correct.  From the point of view of a drug developer hoping to find promising candidates, this may well be the most important statistic, and these the most promising algorithms.

These results compare favorably with previous network-based approaches for drug repurposing.  Perhaps the most successful among the previous approaches is that of Zhang~\etal~\cite{ZYLWLHL18}, who achieved AUROC scores up to 0.87 and AUPR up to 0.26 using their matrix factorization method.  Their method, however, has access to many additional forms of data, such as structure, targets, and enzymes that ours does not, and moreover does not use direct therapeutic interactions as input, so this is not an apples-to-apples comparison.  As discussed in the introduction, we anticipate that in a production setting the methods we study could be combined with methods based on other data to create a hybrid approach that has the best of both worlds.

Also competitive is the approach of Gottlieb~\etal~\cite{GSRS11}, who used an algorithm based on node similarity reminiscent of those discussed in Section~\ref{sec:similarity}, achieving an AUROC score of 0.90, though they give no results for AUPR. Abbas~\etal~\cite{AADNYCCSH21}, testing a large selection of previously published algorithms, achieved generally lower AUROC scores up to 0.70 on their drug-disease network, but impressively good AUPR scores, as high as 0.82 for an algorithm based on SimRank.  The latter should be taken with a pinch of salt, however.  As discussed in Section~\ref{sec:previous}, the network used by Abbas~\etal\ is much denser, by a factor of about~25, than the network we study, and precision is proportional to density, all other factors being equal.  For instance, a no-skill random classifier will guess correct interactions with probability precisely equal to the density.  Thus we would expect a significantly higher AUPR score for any algorithm on the network of Abbas~\etal\ than on a network of much lower density.  Moreover, as mentioned in Section~\ref{sec:previous}, a large majority of the interactions in the network of Abbas~\etal\ appear not to be confirmed drug-disease therapeutic interactions, so the algorithm is for the most part both training on and predicting different kinds of connections than the confirmed interactions that we focus on.  Other studies, such as that of Wang~\etal~\cite{WGWCL15}, do not give quantitative measures of performance against which to make a comparison.

The primary current limitations on our prediction performance are two-fold.  First, algorithms are only as good as the data we feed into them and, while we have taken pains to ensure the quality of our data set as described in Section~\ref{sec:network}, it is limited by the source data from which it was constructed, which is certainly incomplete and may contain errors.  Nonetheless the data are good enough to reliably identify the strongest algorithms for link prediction in this context, and the quality and completeness of the data can be expected to improve over time, so that future studies need only apply those algorithms to such improved data to achieve improved results.  The second limitation of the approach, as mentioned above, is that we use only known therapeutic drug-disease interactions to make our predictions.  As discussed in the introduction, we expect that in a production setting our methods would be combined with other data to improve the quality of the results still further.

One issue we have not tackled in this paper, but which is of potential interest, is the identification of false positives in the data set.  All of our link prediction algorithms give prediction scores for every pair of nodes in the network, both those currently unconnected, which are of interest for repurposing, and those currently connected.  An unusually low score for one of the latter would indicate a drug-disease interaction that is indicated in the data set but which, in the eyes of the algorithm, appears suspicious: if we hadn't already been told of this interaction, we would have been unlikely to predict its existence.  Such drug-disease pairs could be false positives---drugs that do not in fact treat the diseases they are claimed to.  It is a straightforward calculation to find such pairs, but it is not the focus of the present paper so we leave it for future work.

\subsection*{Acknowledgments}
The authors thank Andrew Alt, Carrie Ferrario, Gur-Eyal Sela, and Peter Toogood for useful conversations.  This work was funded in part by the US National Science Foundation under grant numbers DMS--2005899 and DMS--2404617.

\subsection*{Code and data availability}
Data and code for the work described here is available at \texttt{https://github.com/ apolanco115/drug-dis-lp}, with the exception of previously available data and code by other authors, which can be found at the references and locations cited in the text.


\bigskip\noindent\hrulefill\bigskip

\begin{appendices}

\section{Selection of drugs included in the\\data set}
\label{app:data}
As described in Section~\ref{sec:network}, our data set is assembled from a combination of three pre-existing drug databases, DrugBank, NCATS Inxight, and DailyMed.  We start with DrugBank, version 5.1.10, which lists a total of 15\,236 distinct medications.  Not all of these are of therapeutic interest however.  The list also contains things such as cosmetics, foods, allergens, household products, and others, and a crucial step in the preparation of the data is the removal of unwanted entries, which we do in three stages.

First, the DrugBank database assigns each drug to one or more of a set of overlapping status types, including ``approved,'' ``investigational,'' ``withdrawn,'' ``illicit,'' and others.  We restrict ourselves to drugs labeled as ``approved,'' a category that denotes drugs approved for clinical use by the US Food and Drug Administration or one of several comparable agencies in other countries or regions.  This removes 10\,769 drugs from the data set, more than two-thirds of the initial list.

\begin{table}
\centering
\begin{tabular}{lr}
Category & Entries \\
\hline
Allergenic extracts\upstrut & 23 \\
Bacterial toxins & 12 \\
Bee and wasp venom & 12 \\
Coloring agents & 41 \\
Contrast media & 67 \\
Cosmetics & 18 \\
Detergents & 14 \\
Diagnostic agents & 46 \\
Dietary fats & 14 \\
Diet, food, and nutrition & 92 \\
Food additives & 29 \\
Food allergenic extract & 195 \\
Food ingredients & 29 \\
Fungal allergenic extract & 98 \\
Herbs and natural products & 31 \\
Household products & 16 \\
Metals & 72 \\
Mineral supplements & 34 \\
Plant allergenic extract & 62 \\
Pollen allergenic extract & 218 \\
Solvents & 20 \\
Standardized chemical allergens & 54 \\
Sunscreen agents & 27 \\
Sweetening agents & 21 \\
Transition elements & 38 \\
Venoms & 16 \\
\end{tabular}
\caption{Categories of drugs removed from the data set, along with the number in each category.  Note that the total number of drugs removed is less than the sum of the entries in the right-hand column because some drugs belong to more than one category.}
\label{tab:categories}
\end{table}

Second, DrugBank assigns to each drug some number of descriptive labels, including both functional and chemical labels, such as ``amino acids,'' ``anti-bacterial agents,'' or ``analgesics.''  We use these labels to identify entries of minimal therapeutic value, eliminating all those in the categories listed in Table~\ref{tab:categories}.  This removes an additional 866 entries.  There are also some entries in the database that are not listed as belonging to any category.  These ``unknown'' drugs, of which there are 422, we also remove.

These criteria were chosen to focus the network on drugs of established therapeutic value, avoiding non-therapeutic substances.  Including the latter could not only lead to the possibility of their diluting more useful predictions, but could also adversely affect the link prediction process itself, making it less accurate.  By focusing on entries intended for active well-established disease interventions, the network and resulting predictions should better reflect clinically meaningful drug-disease relationships.

In total we end up removing 12\,057 of the starting set of database entries, leaving 3179 core approved drugs in our data set.  A final round of hand cleaning of the data reduces this to the 2620 used for the calculations in the paper.  During the hand cleaning phase, every interaction in the data set was checked manually to confirm its correctness and a range of other housekeeping operations were performed.  This included amalgamating diseases that appeared multiple times in the data set under different names into a single disease.  Moreover, other diseases appear in both generic and specific forms and the drugs associated with these were rationalized so that every drug associated with a specific form is also associated with the generic form (e.g.,~all lung cancer drugs are, by definition, also cancer drugs).  In addition, some simple text parsing errors were fixed in the hand cleaning stage, as were (rare) cases of AI ``hallucination,'' in which the LLM found an indication for a drug that was not supported by the original data.

\section{Theoretical limits on performance}
\label{app:bounds}
Missing data and measurement error can limit the performance of link prediction algorithms in cross-validation tests on real-world data.  Even an algorithm that can predict missing edges perfectly will \emph{appear} to fail if it correctly predicts an edge that is missing from the original data set---an action that will not be credited as a correct prediction in the cross-validation setting.  As we show in this appendix, these effects place limits on the maximum value the AUROC score can attain.

Suppose out of all node pairs that are not connected by an edge in our data set, a fraction~$\mu$ are in fact connected in reality, but those edges are missing from the data set for some reason---because of experimental error or simply because no measurement of them has ever been made.  The remaining fraction $1-\mu$ are genuinely not connected by an edge, either in the data set or in reality.

Now we run a cross-validation experiment on the network using some link prediction algorithm.  A certain fraction of the observed edges are removed and we attempt to predict them.  The link prediction algorithm produces a list of node pairs, drawn from those that are unconnected in the training set, in order from most likely predictions to least likely.  Suppose we take the first $k$ entries from the top of this list as our edge predictions.  For any value of~$k$, let us define $p$ to be the true positive rate for this~$k$, where we include as ``positives'' both those edges that were removed for cross-validation and those that were erroneously missing from the data from the outset.  And let $q$ to be the corresponding false-positive rate for this value of~$k$.

As we let the value of $k$ vary from zero up to the entirety of the list, the values of $p$ and $q$ describe an ROC curve, but unfortunately this is not a curve we can actually measure, because we do not know the identities of all the edges missing from the data set.  We know some of them---the ones that we ourselves removed for cross-validation---but not the ones that were missing from the outset.  Still, if we did somehow know all the missing edges then we could calculate the area~$A_0$ under the curve from
\begin{equation}
A_0 = \int_0^1 p \>\dd q.
\label{eq:A0}
\end{equation}

Since we do not know the identity of the edges that are missing from the outset, any link predictions that identify these edges will be wrongly labeled in our cross-validation tests as false positives, when really they are true positives.  If we make the assumption that our algorithm is equally good at predicting both the edges removed for cross-validation and edges missing from the outset, then this implies that the estimated true positive rate, as calculated over the edges removed for cross-validation, will remain unchanged at~$p$, but the false positive rate will be exaggerated and will now take on the larger value
\begin{equation}
q' = \mu p + (1-\mu) q.
\label{eq:qprime}
\end{equation}
Here the term $\mu p$ represents the genuine missing edges that are predicted to be present with probability~$p$ but then wrongly labeled as false positives, while the $(1-\mu)q$ represents actual false positives---non-edges that are falsely predicted with probability~$q$.

It is worth pausing for a moment over our assumption that the algorithm is equally good at predicting both types of edges.  While this is a reasonable baseline assumption, it could be violated under some circumstances.  For instance, it could be that the edges missing from the data set are missing precisely because they are harder to predict: perhaps all the easy-to-find edges have already been added to the data set and those left in the ``missing'' set are the ones for which link prediction works poorly.  Here we assume this not to be the case, so that the average probability of all true positive predictions takes the same value~$p$ as it does for the edges removed for cross-validation.

With this assumption, the AUROC value that we calculate in our cross-validation experiment is
\begin{equation}
A = \int_0^1 p \>\dd q'
  = \mu \int_0^1 p \>\dd p + (1-\mu) \int_0^1 p \>\dd q
  = \tfrac12 \mu + (1-\mu)A_0,
\end{equation}
where we have used Eq.~\eqref{eq:A0} in the last equality.  The largest possible value of $A_0$, if we had a perfect prediction algorithm, would be~1.  Hence an upper bound on the actual measured area under the curve is
\begin{equation}
A \le 1 - \tfrac12 \mu.
\label{eq:muineq}
\end{equation}
Thus, for example, if 10\% of true edges are missing from the data set, it will be impossible for any algorithm to achieve an AUROC of greater than $1 - \frac12 \times0.1 = 0.95$.  Alternatively, we can reverse the argument and say that if we observe an AUROC of $A$ then the fraction~$\mu$ of edges that could be missing from the data set satisfies
\begin{equation}
\mu \le 2 (1-A).
\label{eq:bound1}
\end{equation}

This inequality gives us an upper bound on what we can extract from a given network.  It tells us that the fraction~$\mu$ of unobserved edges remaining to be found in the network is always less than $2(1-A)$, where $A$ is the value of the AUROC statistic measured for any link prediction algorithm.  Note that the inequality does not depend on the fraction of edges removed in cross-validation, so one is at liberty to try any fraction one wants, as well as any algorithm one wants, in order to make the AUROC as large as possible and hence obtain the tightest bound.

The inequality satisfies some basic sanity tests.  The smallest value of $A$ is~$\frac12$, which gives $\mu<1$, implying that potentially all edges are true predictions in this case.  This is trivially correct: if all edges were true predictions then even a perfect algorithm would be unable to distinguish between genuinely missing edges and edges removed in cross-validation, so by definition $A=\frac12$.  Conversely, the largest possible value of $A$ is~1, implying perfect success at predicting the removed edges.  This would give us $\mu=0$, meaning that none of our predictions correspond to genuinely missing edges, which again is correct: if your algorithm is 100\% correct at picking out the edges removed in cross-validation, then there can be no other correct predictions diluting the results.

The presence of missing edges in the data also affects the precision, although in a relatively simple way: the number of true positive predictions is reduced by a factor of $1-\mu$ and hence the precision is reduced by the same factor, as are measures proportional to precision such as AUPR and top-$k$ precision.  From~\eqref{eq:muineq} and the fact that $\mu$ is positive we have $2A-1 \le 1-\mu \le 1$, which in practice places relatively tight bounds on $1-\mu$ and hence tells us that precision values are not going to be greatly affected by missing edges.

Returning to AUROC values, suppose now that, in addition to the missing edges in the original data (false negatives) there are also false positives: a fraction~$\nu$ of the observed edges are wrong.  Again we run our cross-validation test, removing some of the edges in the data, including some of these false edges, then attempt to predict the removed edges.  Predictions of the false edges will occur with probability~$q$, not probability~$p$, and hence the true-positive rate estimated in the experiment will be modified to a new value
\begin{equation}
p' = (1-\nu) p + \nu q,
\end{equation}
while the false-positive rate still has the same value as before, Eq.~\eqref{eq:qprime}, so our AUROC score becomes
\begin{align}
A &= \int_0^1 p' \>\dd q' \nonumber\\
  &= \mu(1-\nu) \int_0^1 p \>\dd p + (1-\mu)(1-\nu) \int_0^1 p \>\dd q
     + \mu\nu \int_0^1 q \>\dd p + (1-\mu)\nu \int_0^1 q \>\dd q.
\label{eq:fullA}
\end{align}
Integrating by parts, we have
\begin{equation}
\int_0^1 q \>\dd p = \bigl[ pq \bigr]_{p=0}^1 - \int_0^1 p \>\dd q
  = 1 - A_0,
\end{equation}
and hence~\eqref{eq:fullA} becomes
\begin{align}
A &= \tfrac12 \mu(1-\nu) + (1-\mu)(1-\nu) A_0
     + \mu\nu(1-A_0) + \tfrac12 (1-\mu)\nu \nonumber\\
  &= \tfrac12(\mu+\nu) + [1-(\mu+\nu)] A_0.
\end{align}

We don't know the sign of the coefficient $1-(\mu+\nu)$ in this expression---it could be either positive or negative---but if we assume the fractions of false positives~$\nu$ and false negatives~$\mu$ to be small enough that $\mu+\nu<1$, then the coefficient is positive and setting~$A_0=1$ again gives us an upper bound $A \le 1 - \tfrac12(\mu+\nu)$ and hence
\begin{equation}
\mu + \nu \le 2(1-A),
\end{equation}
which is a generalization of Eq.~\eqref{eq:bound1}.  This expression places a bound on the sum of the fractions of false positives and false negatives in the data.  Since both fractions are positive, it also implies that $\mu \le 2(1-A)$, so our result from before holds even in the presence of false positives in the data.  Moreover we also have $\nu \le 2(1-A)$, so we have an upper bound on the fraction of false positives that could be present.

\section{Expectation-maximization algorithm for PLSA}
\label{app:plsa}
The expectation-maximization (EM) algorithm we use for fitting the PLSA model is a bipartite version of the one proposed in~\cite{BKN11}.  We assume a bipartite network with $m,n$ nodes of types 1 and~2 respectively and a number of edges between node~$u$ of type~1 and node~$v$ of type~2 that follows a Poisson distribution with mean $\vec{r}_u\cdot\vec{s}_v = \sum_{z=1}^K r_{uz} s_{vz}$, where $\vec{r}_u$ and $\vec{s}_v$ are $K$-dimensional vectors with non-negative elements.  The assumption of a Poisson distribution may seem surprising, since our drug-disease network only ever has either zero or one edges between any pair of nodes, but in a sparse setting such as ours there is little difference between a Poisson random variable and a zero-one Bernoulli variable, and the Poisson choice is a practical one that makes the calculations simpler.

There is an ambiguous multiplicative factor between $r_{uz}$ and $s_{vz}$ that prevents them from being fully identifiable: for any set of non-negative numbers~$x_z$ with $z=1\ldots K$, if we multiply $r_{uz}$ by $x_z$ for all~$u$ and divide~$s_{vz}$ by the same quantity for all~$v$, then the expected number of edges between every pair of nodes remains the same, regardless of the values of the~$x_z$.  Here we remove this ambiguity by enforcing the condition
\begin{equation}
\sum_u r_{uz} = \sum_v s_{vz}
\label{eq:normalization}
\end{equation}
for all~$z$.

Our purpose with this model is to fit it to our observed bipartite network and hence extract the values of the $\vec{r}_u$ and $\vec{s}_v$, which we can use to estimate the probability of an edge between any two nodes.  We perform the fit by the method of maximum-likelihood.  Let $\mat{B}$ be the $m\times n$ incidence matrix of the network with elements~$B_{uv}=1$ if there is an edge between nodes $u$ and $v$ and 0 otherwise.  Then the likelihood of the network is a product of Poisson distributions:
\begin{equation}
P(\mat{B}|\vec{r},\vec{s}) = \prod_{u=1}^m\>\prod_{v=1}^n
   {( \vec{r}_u\cdot\vec{s}_v )^{B_{uv}}\over
   B_{uv}!} e^{-\vec{r}_u\cdot\vec{s}_v},
\label{eq:likelihood}
\end{equation}
and the log-likelihood is
\begin{equation}
\log P(\mat{B}|\vec{r},\vec{s})
  = \sum_{uv} \biggl[ B_{uv} \log \sum_{z=1}^K r_{uz} s_{vz}
    - \sum_{z=1}^K r_{uz} s_{vz} \biggr],
\label{eq:loglike}
\end{equation}
where we have written out the dot products in full and made use of the fact that $B_{uv} = 0$ or~1 so that $B_{uv}!=1$ for all $u,v$.  Now we employ Jensen's inequality in the form $\log \sum_i x_i \ge \sum_i q_i \log(x_i/q_i)$, where the $x_i$ are any set of positive reals and the $q_i$ are any set of positive reals that sum to~1.  Applying this inequality to the first term in~\eqref{eq:loglike} we get
\begin{equation}
\log P(\mat{B}|\vec{r},\vec{s})
   \ge \sum_{uvz} \biggl[ B_{uv} q_{uv}(z) \log {r_{uz} s_{vz}\over q_{uv(z)}}
    - r_{uz} s_{vz} \biggr],
\label{eq:freeneergy}
\end{equation}
where $q_{uv}(z)$ is any set of positive quantities such that $\sum_z q_{uv}(z) = 1$.

The exact equality is achieved when
\begin{equation}
q_{uv}(z) = {r_{uz} s_{vz}\over\sum_z r_{uz} s_{vz}},
\label{eq:estep}
\end{equation}
meaning also that this choice maximizes the right-hand side of~\eqref{eq:freeneergy} with respect to~$q_{uv}(z)$.  Thus by maximizing with respect to~$q_{uv}(z)$ and then maximizing the resulting expression with respect to $\vec{r}_u$ and $\vec{s}_v$ for all $u,v$ we obtain the maximum-likelihood solution we seek.  To put that another way, a double maximization of the right-hand side of~\eqref{eq:freeneergy} with respect both to $q_{uv}(z)$ and to $\vec{r}_u, \vec{s}_v$ will achieve our goals.  In the EM algorithm we perform this double maximization by simply maximizing repeatedly and alternately with respect to $q_{uv}(z)$ using Eq.~\eqref{eq:estep} and with respect to $\vec{r}_u, \vec{s}_v$ by differentiation.

Differentiating the right-hand side of~\eqref{eq:freeneergy} with respect to $r_{uz}$ and $s_{vz}$ for fixed~$q_{uv}(z)$ gives us the two equations
\begin{equation}
r_{uz} = {\sum_v B_{uv} q_{uv}(z)\over\sum_v s_{vz}}, \qquad
s_{vz} = {\sum_u B_{uv} q_{uv}(z)\over\sum_u r_{uz}}.
\label{eq:rs}
\end{equation}
Summing the first of these over~$u$ and using Eq.~\eqref{eq:normalization} gives
\begin{equation}
\sum_u r_{uz} \sum_v s_{vz} = \biggl( \sum_u r_{uz} \biggr)^2
  = \biggl( \sum_v s_{vz} \biggr)^2
  = \sum_{uv} B_{uv} q_{uv}(z),
\end{equation}
and hence
\begin{equation}
\sum_u r_{uz} = \sum_v s_{vz} = \sqrt{\textstyle{\sum_{uv} B_{uv} q_{uv}(z)}}.
\end{equation}
Then \eqref{eq:rs} becomes
\begin{equation}
r_{uz} = {\sum_v B_{uv} q_{uv}(z)\over\sqrt{\sum_{uv} B_{uv} q_{uv}(z)}},
\qquad
s_{vz} = {\sum_u \bigstrut B_{uv} q_{uv}(z)\over\sqrt{\sum_{uv} B_{uv} q_{uv}(z)}}.
\label{eq:mstep}
\end{equation}
Note that because $q_{uv}(z)$ always appears in the combination $B_{uv} q_{uv}(z)$ it is only necessary to calculate it for node pairs~$u,v$ that are joined by an edge in the network.  Values for all other $u,v$ contribute zero to the sums.

We now have a complete algorithm for estimating $r_{uz}$ and~$s_{vz}$.  Starting from an initial guess at their values (such as random numbers), we compute $q_{uv}(z)$ from Eq.~\eqref{eq:estep} for all $z$ and all $u,v$ connected by an edge, then we calculate updated values of $r_{uz}$ and~$s_{vz}$ from Eq.~\eqref{eq:mstep}, and repeat the whole process until convergence is achieved.  If there are $M$ edges in total in the network then the running time is $\Ord(KM)$ per round of the algorithm, and total running time is this amount times the number of rounds.

Finally, to perform link prediction, we calculate a prediction score within the fitted model for each possible node pair~$u,v$ as the expected number of edges $\sum_z r_{uz} s_{vz}$ between those nodes, and generate edge predictions in order from highest score to lowest.

\section{Algorithm parameters}
The behavior of some of the algorithms we study is controlled by various user-determined parameters, listed in Table~\ref{tab:hyperparams_summary} along with the values used in our calculations.  These values were in general chosen by performing a coarse search over the parameter space to optimize prediction performance according to the AUPR measure.   Some parameters were found to have a negligible impact on performance and these were left at their default or canonical values.  Across all algorithms, the most parameter whose variation had the greatest effect was the dimension of the learned embedding, which has a substantial effect on AUPR values.

\begin{table}
\centering
\begin{tabularx}{\textwidth}{lX}
Algorithm & Parameters \\
\midrule
SVD      & \verb|embedding dimension| = 60\\[4pt]
PLSA     & \verb|embedding dimension| = 90 \\[4pt]
NNMF     & \verb|embedding dimension| = 80\\[4pt]
Node2vec & \verb|embedding_dim| = 256, \verb|walk_length| = 20, \verb|context_size| = 20, \verb|walks_per_node| = 40, \verb|batch_size| = 128, \verb|num_epochs| = 200, \verb|lr = 0.01|, \verb|p| = 1.0, \verb|q| = 1.0 \\[4pt]
GEBE\textsuperscript{p} & \verb|embedding size| = 96, \verb|heat constant| = 1 \\[4pt]
BPR & \verb|no_components| = 100, \verb|num_epochs| = 650, \verb|learning_rate| = 0.01 \\[4pt]
ICTC & \verb|model1| = LGAE, \verb|model2| = GAE, \verb|learning_rate| = 0.01, \verb|num_epoch| = 200, \verb|hidden1_dim| = 32, \verb|hidden2_dim| = 16, \verb|numexp| = 10 \\
\end{tabularx}
\caption{Parameters controlling each algorithm.}
\label{tab:hyperparams_summary}
\end{table}

\end{appendices}

\end{document}